\documentclass[aps,prb,reprint,superscriptaddress]{revtex4-2}
\usepackage{graphicx}	
\usepackage{siunitx}

\usepackage{xcolor}
\usepackage{amsmath}
\usepackage{amssymb}
\usepackage{hyperref}
\usepackage[all]{hypcap} 
\hypersetup{
	colorlinks=false,
	linkcolor=blue,
	urlcolor=blue,
	citecolor=blue,
	pdfpagemode=FullScreen,
}
\usepackage{printlen}
\usepackage[capitalise]{cleveref}

\usepackage{enumitem}
\usepackage{float}
\newcounter{myfigure}
\makeatletter
\def\maketitle{
	\@author@finish
	\title@column\titleblock@produce
	\suppressfloats[t]}	
\@addtoreset{figure}{myfigure}
\makeatother

\usepackage{soul}

\begin{document}
\title{Demonstration of a Field-Effect Three-Terminal Electronic Device with an Electron Mobility Exceeding 40 Million cm$^2$/(Vs)}
\author{T. J. Martz-Oberlander}
\affiliation{Department of Physics, McGill University, Montr\'eal, Qc, H3A 2T8, Canada}
\author{B. Bulgaru}
\affiliation{Department of Physics, McGill University, Montr\'eal, Qc, H3A 2T8, Canada}
\author{Z. Berkson-Korenberg}
\affiliation{Department of Physics, McGill University, Montr\'eal, Qc, H3A 2T8, Canada}
\author{Q. Hawkins}
\affiliation{Department of Physics, McGill University, Montr\'eal, Qc, H3A 2T8, Canada}
\author{K.W. West}
\affiliation{Department of Electrical Engineering, Princeton University, Princeton NJ 08544 USA}
\author{K.W. Baldwin}
\affiliation{Department of Electrical Engineering, Princeton University, Princeton NJ 08544 USA}
\author{A. Gupta}
\affiliation{Department of Electrical Engineering, Princeton University, Princeton NJ 08544 USA}
\author{L. N. Pfeiffer}
\affiliation{Department of Electrical Engineering, Princeton University, Princeton NJ 08544 USA}
\author{G. Gervais}
\email{gervais@physics.mcgill.ca}
\affiliation{Department of Physics, McGill University, Montr\'eal, Qc, H3A 2T8, Canada}

\begin{abstract} 
We report the fabrication and operation of a source-drain-gate three-terminal field-effect electronic device with an electron mobility exceeding $40\times 10^6$ cm$^2$ / (Vs). Several devices were fabricated, with the highest achieved electron mobility obtained using a symmetrically-doped GaAs/AlGaAs quantum well forming a two-dimensional electron gas (2DEG) with a density of $1.47(1) \times 10^{11}$ cm$^{-2}$ and a pristine, pre-fabrication electron mobility of $44(2) \times 10^6$ cm$^2$/(\text{Vs}). To circumvent the well-known degradation of electron mobility during fabrication, devices were fabricated using a flip-chip technique where all lithographic processing steps were performed on a separate sapphire substrate. This method demonstrates the successful operation of various gate assembly designs on distinct 2DEGs without observable mobility degradation. This advance doubles the previous record for field-effect electronic device mobility and enables access to new regimes of quantum transport and applications that were previously unfathomable due to mobility limitations.
\end{abstract}

\date{\today}
\pagenumbering{arabic}

\maketitle

\setlength{\parskip}{1em}

\par \textit{Introduction.} Ever since the development of molecular beam epitaxy (MBE) and modulation doping \cite{Dingle1978} in semiconducting materials, the electron mobility in two-dimensional electron gases (2DEGs) has steadily increased, leading to spectacular discoveries and technological developments. For instance, as electron mobility reached new heights in GaAs/AlGaAs heterostructures, it enabled the development of high-electron-mobility transistors (HEMTs) \cite{Mimura1979,Mimura1980}, which found applications as low-noise RF amplifiers for consumer electronics. Subsequently, high-mobility 2DEGs based on the same material allowed for the observation of the fractional quantum Hall effect (FQH) \cite{Tsui1982} -- an {\it a priori} unexpected quantum liquid of electrons exhibiting fractionalized excitation charges, quantum numbers, and statistics. In subsequent decades, significant progress in GaAs/AlGaAs growth by MBE generated a series of advances in two-dimensional quantum physics and established the exciting possibility of using the FQH effect as a basis for fault-tolerant topological quantum computations \cite{Nayak2008}.

For many years, the electron mobility in the cleanest GaAs/AlGaAs heterostructures plateaued in the $\sim $30$\text{--}33 \times 10^6$ cm$^2$/(Vs) range. While several important discoveries were made on pristine materials (see, e.g., \cite{Yang2022}), fabricating ultra-high electron mobility field-effect devices remained challenging. Due to various factors detrimental to 2DEG quality, and within one spectacular exception recently achieved  in graphene \cite{Geim2025}, field-effect-gated devices exhibited post-fabrication mobilities of no more than $20\times10^{6}$ cm$^{2}$/(Vs) \cite{miller2007} despite having access to higher-mobility materials. Very recently, an important advance in GaAs/AlGaAs MBE growth and engineering led to 2DEGs with electron mobility reaching $57\times10^{6}$ cm$^{2}$/(Vs) \cite{chung2022}, breaking the previous world record significantly. Nevertheless, the persistent challenge of mobility degradation during fabrication remains a critical bottleneck. This work presents a solution: a flip-chip fabrication method in which the GaAs/AlGaAs wafer undergoes no processing. Here, we demonstrate a source-drain-gate three-terminal field-effect device with electron mobility exceeding $40\times10^{6}$ cm$^{2}$/(Vs), with no measurable degradation from the pristine material.

\begin{figure}[th!]
	\centering
	\includegraphics[width=.85\linewidth]{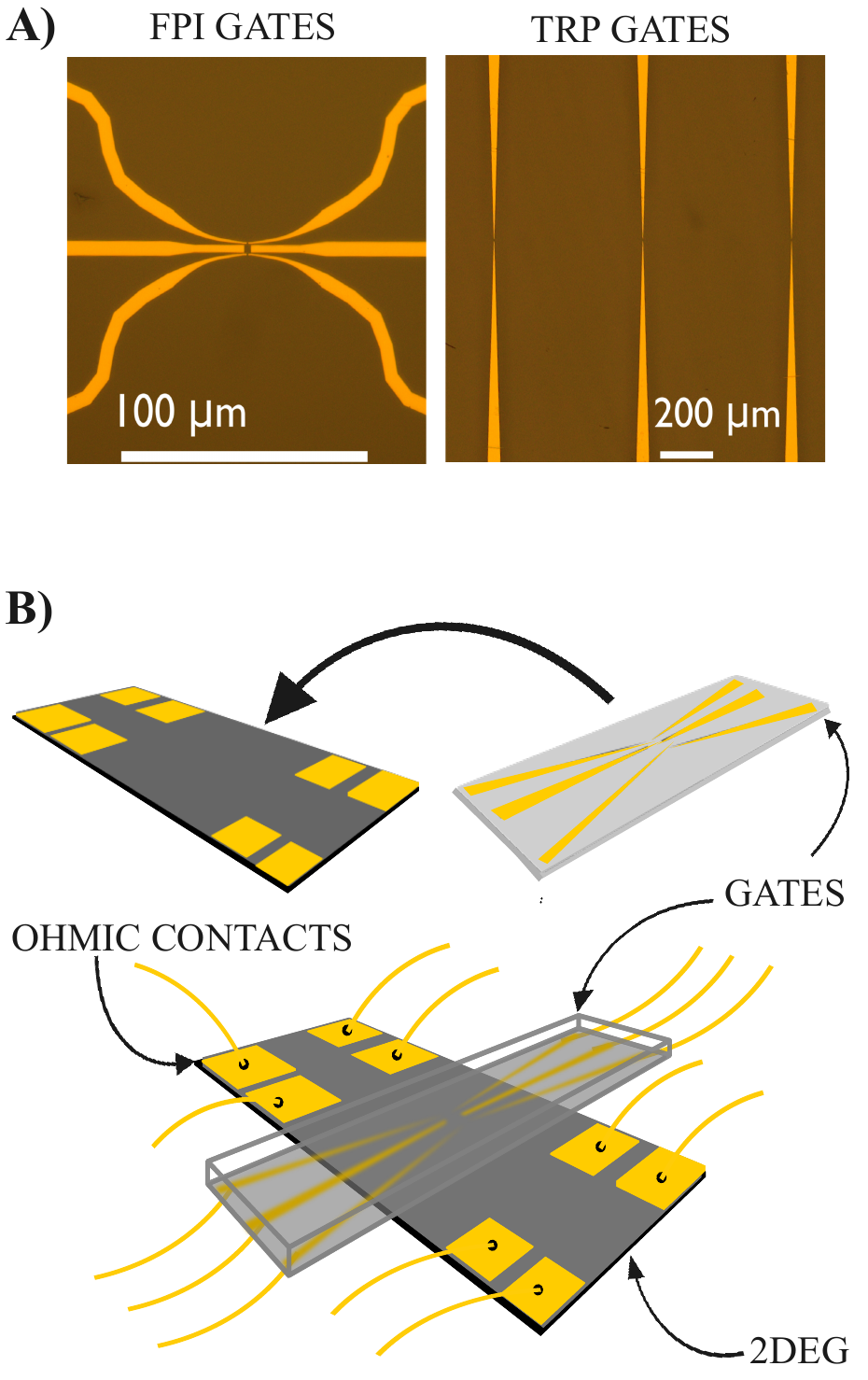}
	\caption{\textbf{{\bf Design of the field-effect gate assembly and representation of the flip-chip fabrication process}. (A)} Left panel: optical image of gates fabricated using electron beam lithography (EBL). This gate assembly, denoted FPI, consists of a set of plunger gates with a 2 $\mu$m tip separation, positioned between two quantum point contacts (QPCs) with a 400 nm tip separation. Right panel: optical image of a different gate assembly design fabricated with photolithography.  This gate assembly, denoted TRP, consists of three parallel QPCs separated by 500 $\mu$m, each with a 3.5 $\mu$m tip separation.  \textbf{(B)} Device fabrication schematic showing that metallic gates were fabricated on a sapphire substrate using EBL prior to mechanical assembly with the GaAs/AlGaAs wafer hosting the 2DEG. The wafer itself is unprocessed except for ohmic contacts formed by diffusion.}
	\label{fig:1}
\end{figure}

\par \textit{Flip-chip electronic device approach.} 
As reported in our earlier work \cite{Bennaceur2015} (see also Ref.~\cite{Beukman2015}), the flip-chip-based electronic field-effect device consists of two independent components: the GaAs/AlGaAs wafer hosting the 2DEG and the gate assembly patterned on a sapphire substrate, as shown in Fig. \ref{fig:1}. Fabricating the gate assembly separately from the semiconductor 2DEG host material offers several advantages: ({\it i}) it eliminates any chemical processing of the wafer, ({\it ii}) it minimizes or removes possible strain on the GaAs/AlGaAs crystalline structure due to differential thermal contraction from the metallic gates and the substrate during cooling to low temperatures, and ({\it iii}) the GaAs/AlGaAs heterostructure is never exposed to X-rays generated during the electron beam processing required to fabricate sub-micron features. In conventional direct fabrication methods, each of these factors is known to irreversibly decrease the electron mobility to some extent. This damage can degrade device quality and performance, causing the disappearance of the most fragile quantum phases that only occur in the highest electron mobility 2DEGs under magnetic fields at low temperatures, thereby hindering their study {\it via} field-effect gating. \\ 

The flip-chip process also provides significant practical advantages compared to conventional direct processing approaches. Namely, it prevents wasting ultra-high electron mobility GaAs/AlGaAs material due to irreversible fabrication errors (such as malfunctioning gates) since the entire device can be disassembled and re-fabricated with a new gate assembly. Additionally, it allows the same 2DEG host material to be tested and characterized with different field-effect gate configurations, enabling the study of distinct devices such as transistors \cite{Mimura1980}, quantum point contacts (QPCs) \cite{miller2007}, and complex interferometers \cite{willett2013,fu2016,Dutta2021,Dutta2022}, to name a few. \\

\par \textit{Device fabrication.} Devices and measurements were made on two distinct GaAs/AlGaAs 2DEG heterostructures grown by MBE and two different gate assembly designs. Data from three distinct devices are shown in this work (see Fig.~\ref{fig:3}).  The heterostructure details for both wafers are provided in the Supplemental Material (SM), where we identify them as labeled by the MBE growth team.  The 2DEG in wafer numbered \#3-11-10.2 has an electron density of $3.1(1) \times 10^{11}$ cm$^{-2}$ and electron mobility ranging from $25\text{--}30\times10^{6}$ cm$^{2}$ / (Vs) depending on cooldown conditions with a red LED illumination down to $\sim$5 K.  The 2DEG hosted in wafer \#P5-4-21.1 has an electron density of $1.47(1) \times 10^{11}$ cm$^{-2}$ and electron mobility ranging from $40\text{--}46\times 10^6$ cm$^2$/(Vs), also depending on cooldown conditions. From the raw wafers, smaller pieces were cleaved into rectangles measuring $\sim$$1.4\times 8$ mm. Ge/Ni/Au ohmic contacts were then deposited and annealed at their corners, then affixed to a fibreglass (G10) header using 25 micron gold wire and indium solder.\\

Electronic gate assemblies were fabricated on separate sapphire (Al$_2$O$_3$) substrates using photolithography for the TRP device (see Fig.~\ref{fig:1}{\bf A}, right panel) and EBL for the FPI device (see Fig.~\ref{fig:1}{\bf A}, left panel). In both cases, 150/7 nm of Au/Ti was evaporated on the sapphire substrate, which was then cut into $\sim$$2 \times 8$ mm pieces. Two distinct TRP devices were fabricated; one included a 40 nm Al$_2$O$_3$ layer deposited by atomic layer deposition (ALD) while the other did not. Final device assembly proceeded as follows (see Fig.~\ref{fig:1}). First, surface irregularities (such as oval defects \cite{Chand1990} known to exist on the GaAs/AlGaAs) were identified under an optical microscope to determine the most defect-free area. The gate assembly was then ``flipped'' over the wafer with gates facing the GaAs surface, and 50 micron gold wires were affixed from the gates to the G10 header. Finally, as described in our previous work \cite{Bennaceur2015}, the entire device was mechanically assembled and fastened using BeCu springs and stainless steel screws, applying slight pressure on the gate assembly toward the GaAs wafer for mechanical stability.  \\

\begin{figure}[tbh!]
	\centering
	\includegraphics[width=1.\linewidth]{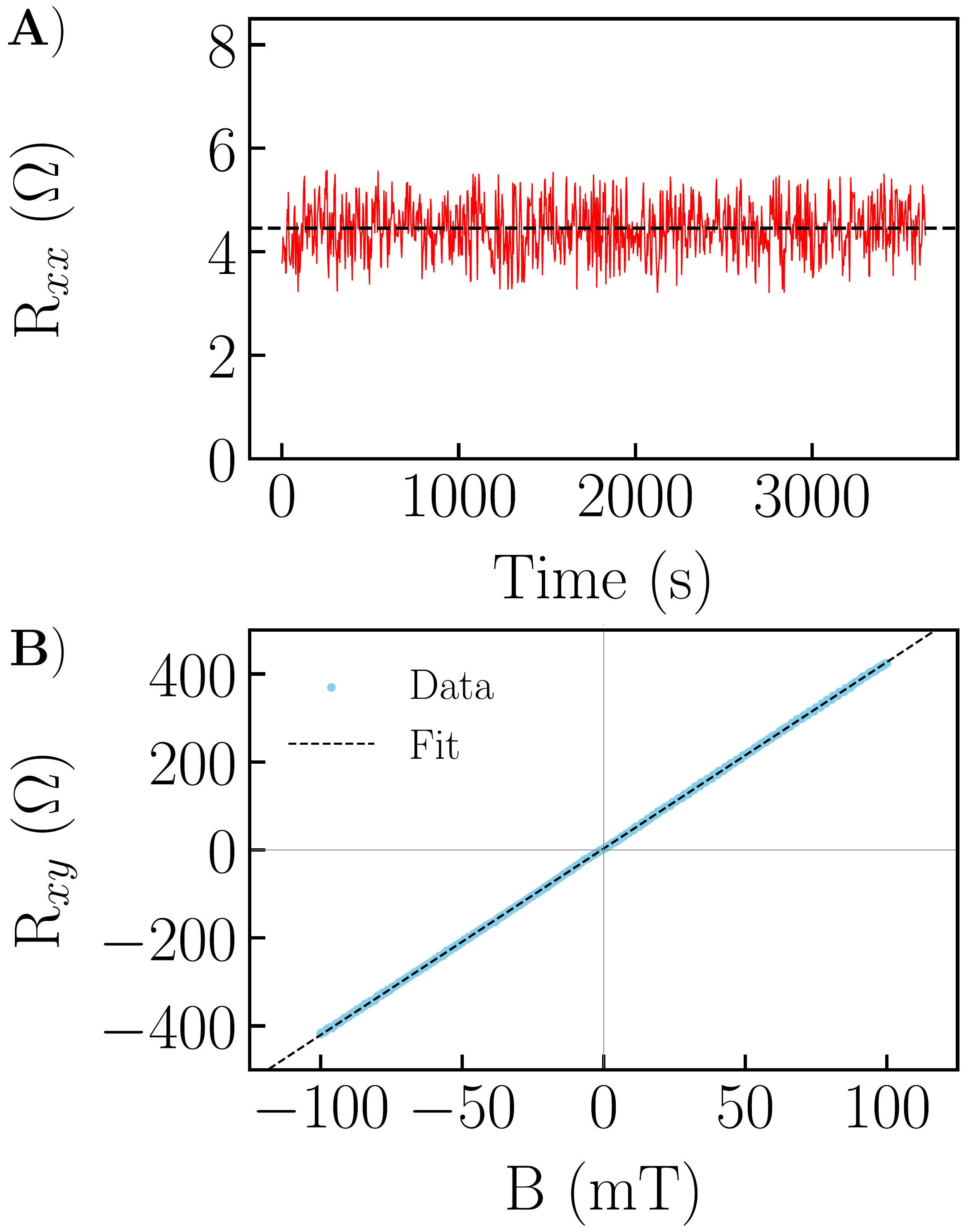}
	\caption{\textbf{{\bf Electron density and mobility measurements for the highest electron mobility flip-chip device fabricated}. (A)} Longitudinal resistance $R_{xx}$ of the flip-chip device with the FPI gate assembly (see Fig.~\ref{fig:1}) mounted on GaAs/AlGaAs wafer \#P5-4-21.1, measured at 14 mK versus time. The dotted line shows the computed average with a filter of $R_{xx} = 4.45(10)$ $\Omega$.  \textbf{(B)} Hall resistance $R_{xy}$ of the same device versus magnetic field. The electron density was extracted from the linear fit (dotted line), yielding a value of $1.47(1)\times 10^{11}$ cm$^{-2}$, corresponding to an electron mobility of $44(2)\times 10^6$ cm$^2$/(Vs). This mobility is in excellent agreement with prior measurements at Princeton by the GaAs/AlGaAs growth team at $\sim$300 mK.\\}
	\label{fig:2}
\end{figure}

{\it Basic characterization.} Electronic transport measurements were performed both in an Oxford He$^3$ dry refrigerator (base temperature of $\sim$300 mK) and in an Oxford Triton dilution refrigerator (base temperature of 14 mK). Resistivity was determined using a four-point resistance measurement circuit at a fixed current of 10 nA with an SR830 lock-in amplifier at frequency $f= 54.32$ Hz. 
For the ultra-high mobility 2DEG wafer \#P5-4-21.1 mounted with the FPI gate assembly, the longitudinal resistance $R_{xx}$ was measured at base temperature (14 mK) and averaged over time (see Fig.~\ref{fig:2}{\bf A}). Similarly, Hall resistance $R_{xy}$ was measured across the gated region of the device in a separate measurement at the same temperature, excitation current, frequency, and magnetic field range of $\pm 100$ mT (see Fig.~\ref{fig:2}{\bf B}). From the linear fit (dotted line), electron density $n$ was determined to be $1.47(1) \times 10^{11}$ cm$^{-2}$, yielding an electron mobility $\mu=1/ne\rho$ of $44(2) \times 10^6$ cm$^2$/(Vs), where $e$ is the electric charge and $\rho$ the sheet resistivity. For consistency, the electron mobilities quoted in this work have been measured immediately before any voltage was applied to the gates. We note in passing that the latest generation of GaAs/AlGaAs wafers have varying electron densities and mobilities throughout the 2DEG \cite{haug2019} and we have observed slight variations with time at low temperatures.

\begin{figure}[th!]
	\centering
	\includegraphics[width=.95\linewidth]{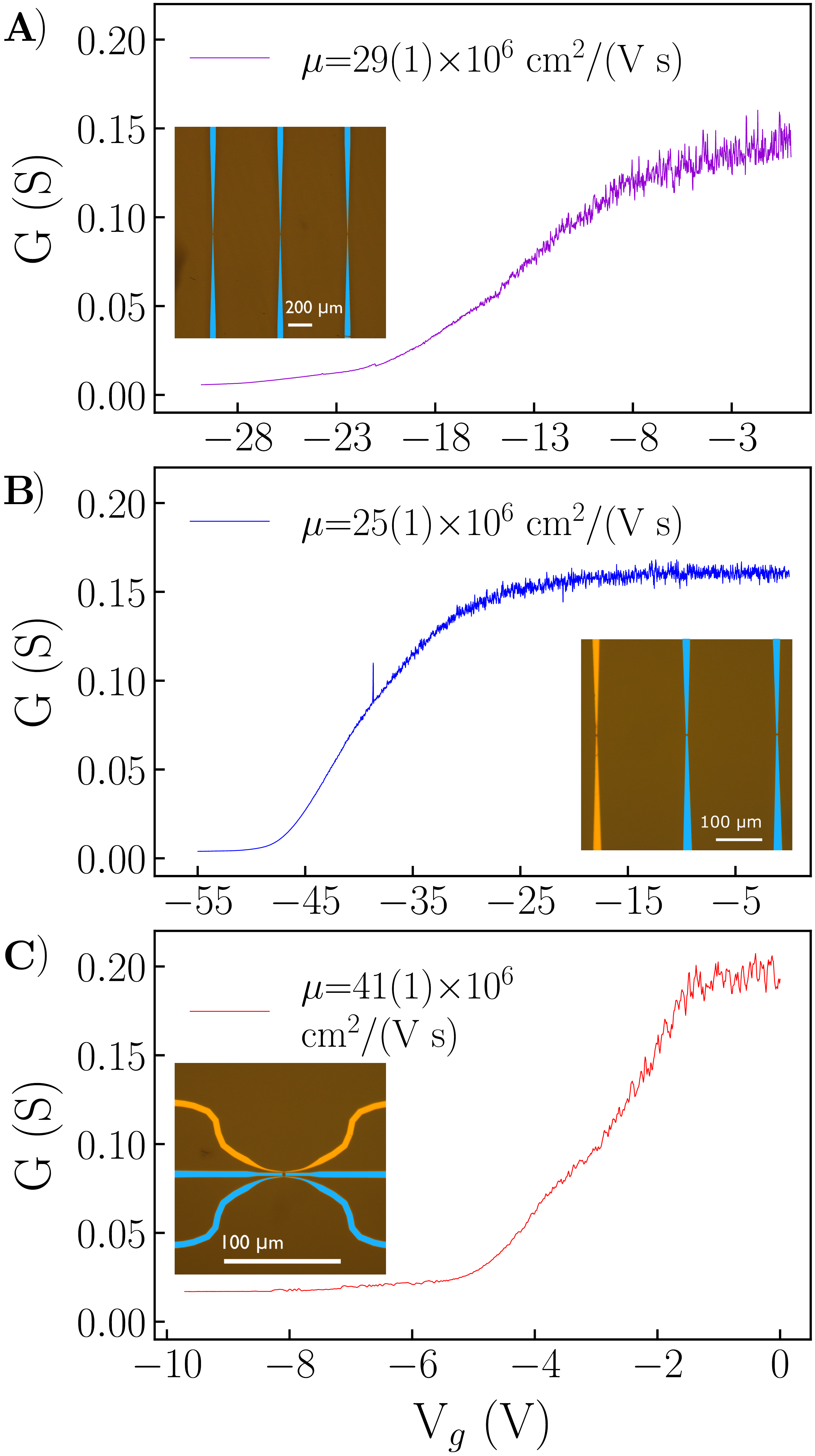}
	\caption{ 
	{\textbf{Conductance versus gate voltage for three distinct flip-chip devices.} {\bf (A)} Conductance $G$ versus gate voltage $V_g$ for a flip-chip device with TRP gates mounted on GaAs/AlGaAs 2DEG wafer \#3-11-10.2, with all gates activated at the same voltage $V_g$. {\bf (B)} Conductance $G$ versus gate voltage $V_g$ for a second device based on the same TRP gate design mounted on a different piece from the same wafer, with a 40 nm thick Al$_2$O$_3$ layer deposited {\it via} ALD as a final capping layer, with four out of six gates activated at the same voltage $V_g$ (shown by blue false coloring). {\bf (C)} Conductance $G$ versus gate voltage $V_g$ for a flip-chip device with FPI gates mounted on ultra-high mobility GaAs/AlGaAs 2DEG wafer \#P5-4-21.1, with four out of six gates activated at the same voltage $V_g$ (shown by blue false coloring). Optical photographs of the gate assemblies for each measured device are shown in the inset of each panel. In all panels, the electron mobilities measured in the same cooldowns are shown. All data was taken at $\sim$$300$ mK.}}	
	\label{fig:3}
\end{figure}

\par \textit{Three-terminal conductance measurements.} The conductance $G$ versus gate voltage $V_g$ for two representative flip-chip devices is shown in Fig.~\ref{fig:3}, where panels \textbf{A} and \textbf{B} display measurements for the TRP gate assembly mounted on wafer \#3-11-10.2, while panel \textbf{C} shows data for the FPI gate assembly mounted on wafer \#P5-4-21.1. All data was collected at $\sim$$300$ mK with an AC excitation voltage of 0.14 mV at $f = 54.32$ Hz using an SR830 lock-in amplifier with a voltage divider circuit (see conductance circuit in the SM). 
The current induced {\it via} an applied potential difference across the device and the voltage difference across the 2DEG was measured concomitantly. 
The conductance $G$ is shown versus the unique voltage applied to the gates. The gates used in each measurement are shown in blue false colouring in the inset of each panel.\\

\begin{figure}[th!]
	\centering
	\includegraphics[width=1.\linewidth]{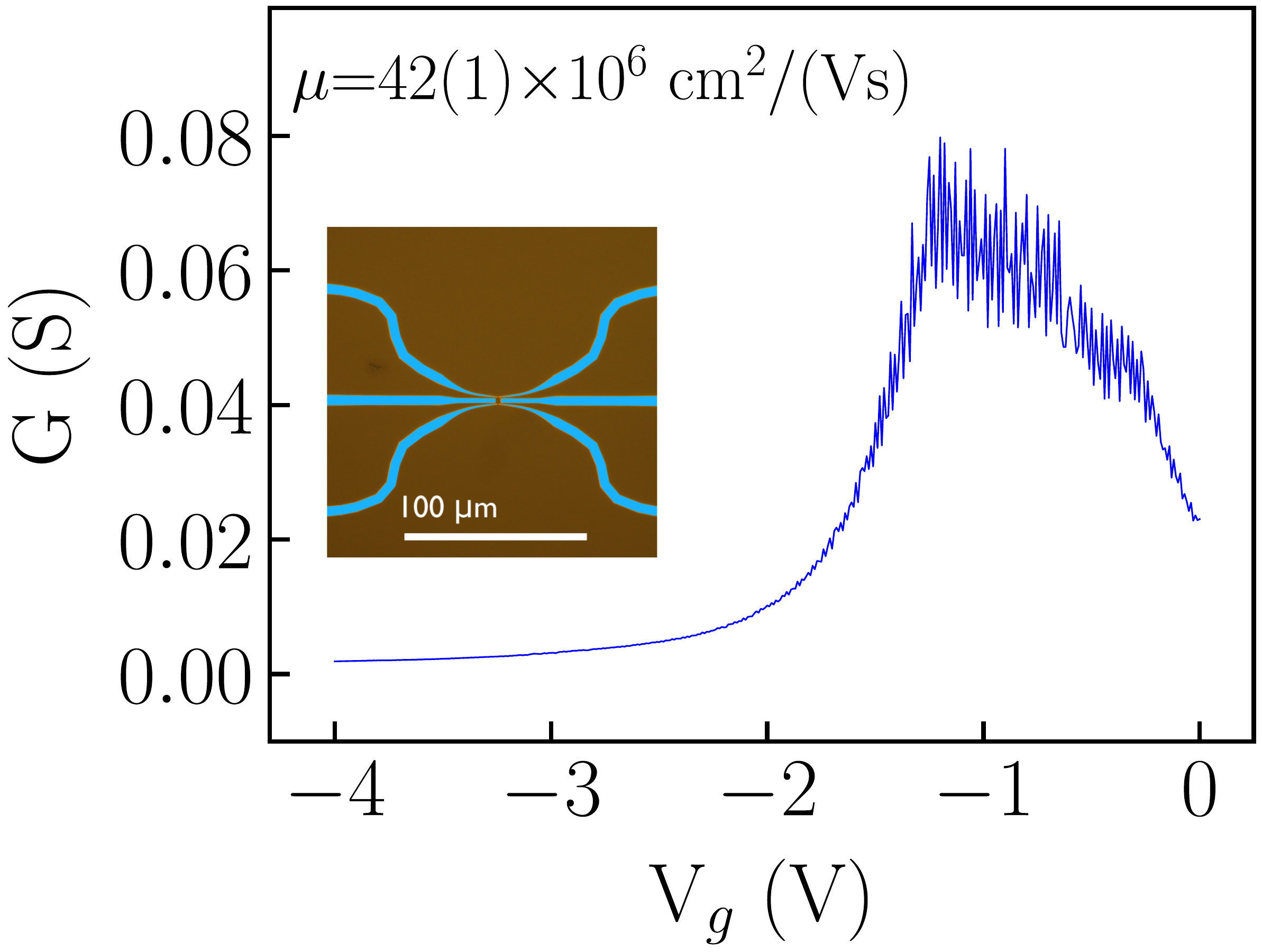}
	\caption{   \textbf{Conductance versus gate voltage in a magnetic field}. Same electronic device as shown in Fig.~\ref{fig:3}{\bf C} (FPI + \#P5.4.21.1) but measured at 14 mK temperature and in a magnetic field $B=1.964$ T, corresponding to a filling factor of $\nu=3.1$. All gates were activated with an equipotential voltage $V_g$ in this measurement, as shown by the blue false coloring of the gates in the inset.}
	\label{fig:4}
\end{figure}

A clear field-effect conductance decrease is observed in all three panels of Fig.~\ref{fig:3} and Fig.~\ref{fig:4} when applying a gate voltage, with distinct threshold voltages ranging from $-2$ V (Fig.~\ref{fig:3}{\bf C}) to higher values such as $-20$ V (Fig.~\ref{fig:3}{\bf B}). Our experience with flip-chip three-terminal devices suggests this variation is not due to differences in gate or heterostructure design but rather to the proximity between the gate assembly and the semiconductor 2DEG surface. For example, remounting a gate assembly on the same wafer with only slight positional changes can modify the conductance pinch-off threshold, likely due to surface defects (such as oval defects) on the GaAs/AlGaAs capping layer. Nonetheless, as shown in Fig.~\ref{fig:3}{\bf C} (and Fig.~\ref{fig:4}), a pinch-off voltage threshold comparable to conventional direct lithography ($\sim$$-2$ V) can be achieved.

The $\sim$$42 \times 10^6$ cm$^2$/(Vs) electron mobility FPI flip-chip device (same as in Fig.~\ref{fig:3}{\bf C}) was also tested at 14 mK in a fixed magnetic field of $B = 1.964$ T, corresponding to a Landau level filling factor $\nu = \frac{n}{(B/\phi_0)}=3.1$ (see Fig.~\ref{fig:4}) where $\phi_0=h/e$ is the flux quantum and  $h$ is the Planck constant.  In this measurement, all gates were activated with an equipotential voltage (blue false coloring in the inset). Similar to Fig.~\ref{fig:3}{\bf C}, a clear conductance pinch-off is observed with a low-voltage threshold. The sharp rise in conductance up to $V_g = -2$ V remains unexplained but may reflect the 2DEG response near the $\nu = 3$ quantum Hall state for this gate design. Regardless, both datasets, with and without magnetic field, unequivocally demonstrate field-effect gating of a 2DEG with electron mobility exceeding $40 \times 10^6$ cm$^2$/(Vs) without degradation compared to the pristine 2DEG material prior to device assembly.\\

\par \textit{Future outlooks and concluding remarks.} We now turn our attention to the potential technological and scientific implications of this work. As previously mentioned, recent advances in MBE growth have produced a new generation of GaAs/AlGaAs 2DEGs with electron mobility reaching $57 \times 10^6$ cm$^2$/(Vs) \cite{chung2022}. 
While not always the definitive figure-of-merit, it is established that increased electron mobility (and reduced disorder) enables new quantum phases and phenomena.
For example, prior work on cleave-edge overgrown 2DEGs revealed spin-charge separation in one-dimensional quantum wires due to Luttinger liquid physics \cite{Yacoby2005}, and the $\nu = 12/5$ fractional quantum Hall state, believed to be a Read-Rezayi state with Fibonacci-sequence-derived quantum statistics \cite{Rezayi2009}. The latter is particularly relevant for fault-tolerant quantum computation as it could serve as a topological qubit platform with universal quantum computing properties \cite{Nayak2008}.
Progress toward this topological quantum computation approach has historically been hindered by the difficulty of preserving such delicate quantum states through the clean-room processing required for device fabrication. Our mechanical flip-chip approach, which maintains pristine electron mobility without degradation and comparable pinch-off thresholds to conventional methods, now demonstrates that this limitation can be overcome.\\

This advance is likely to impact field-effect devices broadly, enabling the fabrication of quantum wires, dot or anti-dot arrays, and electron interferometers with unprecedented ultra-high electron mobility and could extend beyond GaAs/AlGaAs and be applied to other 2DEG materials. In this spirit of fabricating electronic  devices with unprecedented quality, we mention  recent work in graphene \cite{Geim2025} where electron mobility exceeding 50 million cm$^2$/$(\text{Vs)}$ were fabricated using proximity gates, albeit  with  electron density much lower  than in our work,  {\it i.e.}  in the  $\sim 10^{7}-10^8$ cm$^{-2}$ range. Our work could also potentially proved important for  future  materials platforms that could surpass current state-of-the-art 2DEGs in terms of low disorder and device performance.   While engineering studies are ongoing, this method shows promise for low-temperature electronic devices of exceptional quality, including potentially HEMT amplifiers with superior electron mobility -- of interest for the growing field of cryogenic quantum computation. \\

\textit{Acknowledgments.} This work has been supported by NSERC (Canada), PromptInnov/MEI (Qu\'ebec), MITACS, FRQNT-funded strategic clusters INTRIQ (Québec) and Montr\'eal-based CXC. The Princeton University portion of this research is funded in part by the Gordon and Betty Moore Foundation’s EPiQS Initiative, Grant GBMF9615.01 to Loren Pfeiffer. Sample fabrication was carried out at the McGill Nanotools Microfabrication facility, in the clean room  complex of Polytechnique Montr\'eal,  and at the Institut Interdisciplinaire d’Innovation Technologique (3IT) facility at the University of Sherbrooke. We thank M. Abbasi and K. Bennaceur for their contributions to the project, and  S. Ren\'e de Cotret, R. Talbot, R. Gagnon, and J. Smeros for technical assistance. Finally, we thank M.P. Lilly and S.J. Addamane for insightful discussions. \\

\textit{Corresponding author.} Guillaume Gervais, gervais@physics.mcgill.ca.\\

\onecolumngrid
\pagebreak

\begin{center}
	\textbf{\large Supplemental Material: Demonstration of a Field-Effect Three-Terminal Electronic Device with an Electron Mobility Exceeding 40 Million cm$^2$/(Vs)}
\end{center}
%%%%%%%%%% Merge with supplemental materials %%%%%%%%%%
%%%%%%%%%% Prefix a "S" to all equations, figures, tables and reset the counter %%%%%%%%%%
\setcounter{equation}{0}
\stepcounter{myfigure}
\setcounter{table}{0}
\setcounter{page}{1}
\makeatletter
\renewcommand{\theequation}{S\arabic{equation}}
\renewcommand{\thefigure}{S\arabic{figure}}
\renewcommand{\bibnumfmt}[1]{[S#1]}
\renewcommand{\citenumfont}[1]{S#1}

% Make section title lowercase
\def\@hangfrom@section#1#2#3{\@hangfrom{#1#2}#3}
\def\@hangfroms@section#1#2{#1#2}
% Make subfigures capitalized letters
% Make section and subsections numbered
\renewcommand{\thesection}{\arabic{section}}
\renewcommand{\thesubsection}{\thesection.\arabic{subsection}}
%%%%%%%%%% Prefix a "S" to all equations, figures, tables and reset 
\setlength{\parindent}{0pt}          % Removes the indent
\setlength{\parskip}{1em}            % Adds space between paragraphsthe counter
% Make sections and subsections float to the left
\renewcommand\section{%
	\@startsection{section}{1}{\z@}%
	{-3.5ex \@plus -1ex \@minus -.2ex}%
	{2.3ex \@plus.2ex}%
	{\normalfont\large\bfseries\raggedright}%
}
\renewcommand\subsection{%
	\@startsection{subsection}{2}{\z@}%
	{-3.25ex\@plus -1ex \@minus -.2ex}%
	{1.5ex \@plus .2ex}%
	{\normalfont\normalsize\bfseries\raggedright}%
} 
%%%%%%%%%%

\section{Heterostructures}

\begin{figure}[ht]
	\centering % Keep images centered
	% Subfigure A
	\begin{minipage}[t][9.5cm][s]{0.4\linewidth}
		\begin{picture}(30,0)
			\put(-70,0){\raggedright\textbf{A) \#P5-4-21.1}}
		\end{picture}
		\vfill
		\includegraphics[width=\columnwidth]{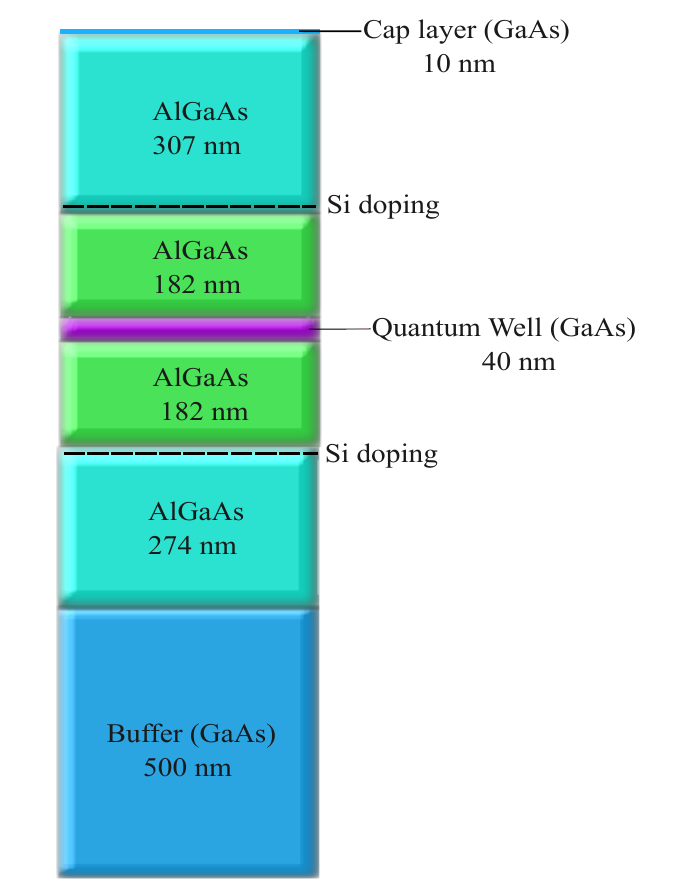}
	\end{minipage}%
	%\hfill % Add horizontal space between them
	% Subfigure B
	%
	\begin{minipage}[t][9.5cm][s]{0.4\linewidth}
		\begin{picture}(30,0)
			\put(-55,0){\raggedright\textbf{B) \#3-11-10.2}}
		\end{picture}
		\vfill
		\includegraphics[width=\columnwidth]{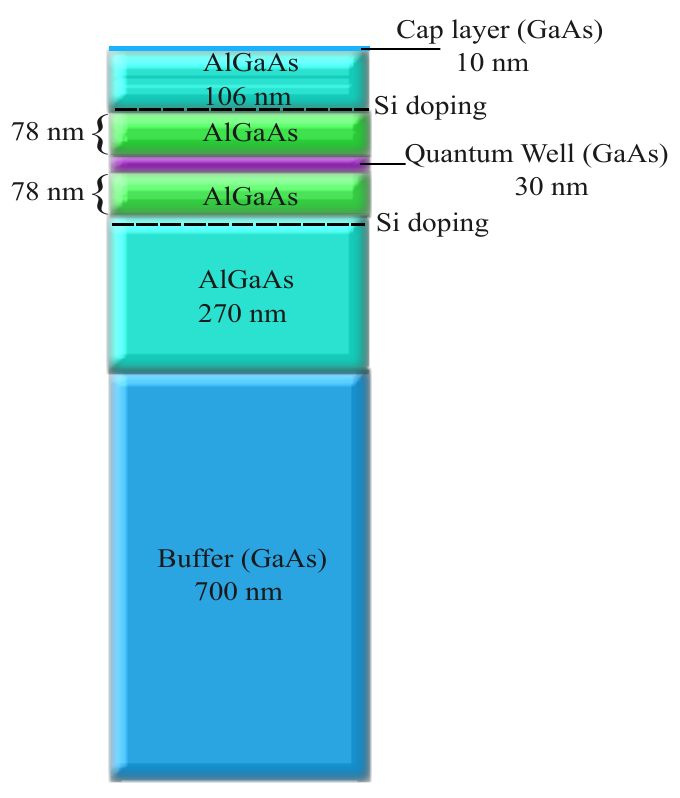}
	\end{minipage}
	
	\caption{Heterostructures for wafer samples \textbf{(A)} \#P5-4-21.1 and \textbf{(B)} \#3-11-10.2, showing the layer structure as grown sequentially by MBE.}
	\label{fig:heterostructures}
\end{figure}

The devices presented in this paper were fabricated using two different wafers, each shown in Fig \ref{fig:heterostructures}. Samples \#P5-4-21.1 and \#3-11-10.2 are grown by molecular beam epitaxy (MBE) on GaAs substrates with thicknesses of 500 nm and 700 nm, respectively. A layer of AlGaAs is grown before the lower doping region to mitigate detrimental effects from the substrate. The separation distance between the doped regions and the quantum well is shown in green. A protective cap layer is added with a composition similar to that of the substrate.

\addvspace{10pt}

\section{Measurement Circuits}

\subsection{Resistance Circuit}

\begin{figure}[ht]
	\centering
	\includegraphics[width=0.7\linewidth]{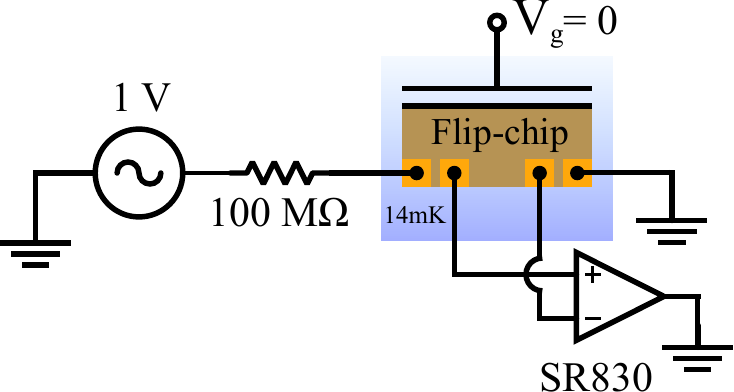}
	\caption{Four-point measurement circuit used to measure resistance across the 2DEG with a fixed current of 10 nA.}
	\label{fig:resistance_circuit}
\end{figure}

The circuit in Fig. \ref{fig:resistance_circuit} illustrates a standard four-point resistance measurement where a fixed current is applied, and the voltage difference across the two-dimensional electron gas (2DEG) is measured. An SR830 lock-in amplifier was used to apply 1.0 V at $f=54.32$ Hz across a 100 M$\Omega$ resistor in series, resulting in a fixed current of 10 nA applied to the flip-chip device.

\subsection{Conductance Circuit}

\begin{figure}[ht]
	\centering
	\includegraphics[width=0.9\linewidth]{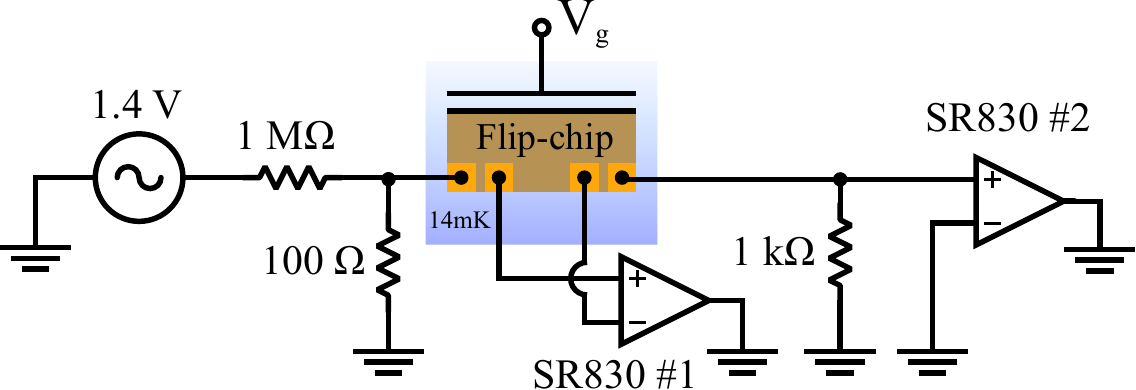}
	\caption{Four-point measurement circuit used to measure conductance across the flip-chip device. The same circuit is used to measure pinch-off of the device by applying a voltage $V_g$ to the gates.}
	\label{fig:conductance_circuit}
\end{figure}

The circuit in Fig. \ref{fig:conductance_circuit} depicts a four-point conductance measurement where a constant voltage is applied across the flip-chip device. The setup includes two synchronized SR830 lock-in amplifiers, a 1 M$\Omega$ $\|$ 100 $\Omega$ voltage divider, and a 1 k$\Omega$ resistor connected in series with the flip-chip device. Gate voltages $V_g$ are applied to distinct gate structures using a KT2400 source meter to operate the gates. Both the voltage difference across the device's ohmic contacts and the current (measured \textit{via} the voltage drop across the 1 k$\Omega$ resistor) are recorded using the lock-in amplifiers.

\section{Resistance Measurement in a Magnetic Field}

\begin{figure}[ht]
	\centering
	\includegraphics[width=0.8\linewidth]{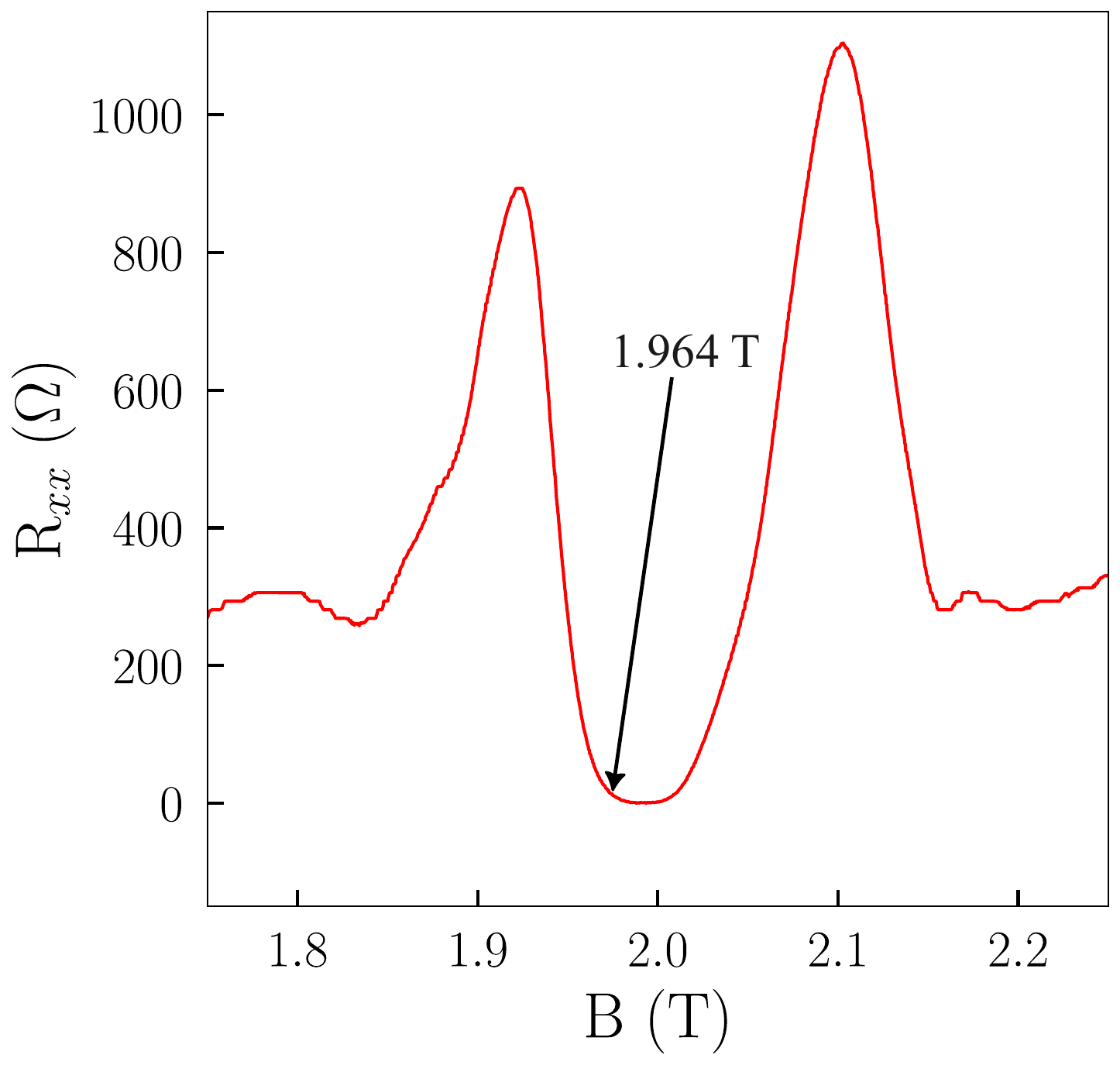}
	\caption{Flip-chip $R_{xx}$ response in the region around $\nu = 3$. The arrow indicates the magnetic field of 1.964 T at which the gate voltage sweep in Fig. \ref{fig:4} of the main text was measured.}
	\label{fig:fieldsweep_zoom}
\end{figure}

The measurement shown in Fig. \ref{fig:4} of the main text presents conductance as a function of gate voltage in a magnetic field of 1.964 T, which is in the vicinity of the Landau level corresponding to a filling factor of $\nu = 3$. Fig. \ref{fig:fieldsweep_zoom} shows the $R_{xx}$ measurement in the neighboring region.

\begin{thebibliography}{0}

 
\bibitem{Dingle1978}
R. Dingle, H. L. Stormer, A. C. Gossard, and W. Wiegmann, Electron mobilities in modulation-doped semiconductor heterojunction superlattices, \href{https://doi.org/10.1063/1.90457}{Appl. Phys. Lett. \textbf{33}, 665 (1978)}.

\bibitem{Mimura1979} 
 T. Mimura, \textit{High Electron Mobility Transistor},
{Japanese Patent JPS5953714B2, filed 1979, published 1984}.


\bibitem{Mimura1980}
T. Mimura, S. Hiyamizu, T. Fujii, and K. Nanbu, A new field-effect transistor with selectively doped GaAs/n-Al$_x$Ga$_{1-x}$As heterojunctions, \href{https://doi.org/10.1143/JJAP.19.L225}{Jpn. J. Appl. Phys. \textbf{19}, L225 (1980)}.

\bibitem{Tsui1982}
D. C. Tsui, H. L. Stormer, and A. C. Gossard, Two-dimensional magnetotransport in the extreme quantum limit, \href{https://doi.org/10.1103/PhysRevLett.48.1559}{Phys. Rev. Lett. \textbf{48}, 1559 (1982)}.

\bibitem{Nayak2008} 
C. Nayak, S. H. Simon, A. Stern, M. Freedman, and S. Das Sarma, Non-Abelian anyons and topological quantum computation, \href{https://journals.aps.org/rmp/abstract/10.1103/RevModPhys.80.1083}{Rev. Mod. Phys. \textbf{80}, 1083 (2008).}

\bibitem{Yang2022}
K. K. W. Ma, M. R. Peterson, V. W. Scarola, and K. Yang, Fractional quantum Hall effect at the filling factor $\nu$ = 5/2, \href{https://doi.org/10.1016/B978-0-323-90800-9.00135-9}{Encycl. Condens. Matter Phys. \textbf{1}, 324 (2024)}.

 \bibitem{Geim2025}
Daniil Domaretskiy, Zefei Wu, Van Huy Nguyen, Ned Hayward, Ian Babich, Xiao Li, Ekaterina Nguyen, Julien Barrier, Kornelia Indykiewicz, Wendong Wang, Roman V. Gorbachev, Na Xin, Kenji Watanabe, Takashi Taniguchi, Lee Hague, Vladimir I. Fal’ko, Irina V. Grigorieva, Leonid A. Ponomarenko, Alexey I. Berdyugin and  Andre K. Geim, Proximity screening greatly enhances electronic quality of graphene, \href{https://doi.org/10.1038/s41586-025-09386-0} {Nature {\bf 644}, 646 (2025).}


\bibitem{miller2007}
J. B. Miller, I. P. Radu, D. M. Zumbühl, E. M. Levenson-Falk, M. A. Kastner, C. M. Marcus, L. N. Pfeiffer, and K. W. West. Fractional quantum Hall effect in a quantum point contact at filling fraction 5/2, \href{https://www.nature.com/articles/nphys658}{Nature Phys. \textbf{3}, 561 (2007).}



\bibitem{chung2022} 
Y. J. Chung, A. Gupta, K. W. Baldwin, K. W. West, M. Shayegan, and L. N. Pfeiffer,
Understanding limits to mobility in ultrahigh-mobility GaAs two-dimensional electron systems: 100 million cm$^2$/Vs and beyond, \href{https://journals.aps.org/prb/abstract/10.1103/PhysRevB.106.075134}{Phys. Rev. B \textbf{106}, 075134 (2022).} 

\bibitem{Bennaceur2015} 
K. Bennaceur, B. A. Schmidt, S. Gaucher, D. Laroche, M. P. Lilly, J. L. Reno, K. W. West, L. N. Pfeiffer, and G. Gervais, Mechanical flip-chip for ultra-high electron mobility devices, \href{https://www.nature.com/articles/srep13494}{Sci. Rep. \textbf{5}, 13494 (2015).}

\bibitem{Beukman2015}
A. J. A. Beukman, F. Qu, K. W. West, L. N. Pfeiffer, and L. P. Kouwenhoven, A noninvasive method for nanoscale electrostatic gating of pristine materials, \href{https://doi.org/10.1021/acs.nanolett.5b02800}{Nano Lett. \textbf{15}, 6883 (2015).}

\bibitem{willett2013}
R. L. Willett, C. Nayak, K. Shtengel, L. N. Pfeiffer, and K. W. West, Magnetic-field-tuned Aharonov-Bohm oscillations and evidence for non-Abelian anyons at $\nu = 5/2$, \href{https://doi.org/10.1103/PhysRevLett.111.186401}{Phys. Rev. Lett. \textbf{111}, 186401 (2013).}

\bibitem{fu2016}
H. Fu, P. Wang, P. Shan, L. Xiong, L. N. Pfeiffer, K. W. West, M. A. Kastner, and X. Lin, Competing $\nu = 5/2$ fractional quantum Hall states in confined geometry, \href{https://doi.org/10.1073/pnas.1614543113}{Proc. Natl. Acad. Sci. U.S.A. \textbf{113}, 12386 (2016).}

\bibitem{Dutta2021}  
B. Dutta, W. Yang, R. A. Melcer, H.~K. Kundu, M. Heiblum, V. Umansky, Y. Oreg, A. Stern, and D. Mross, Distinguishing between non-Abelian topological orders in a quantum Hall system,
\href{https://www.science.org/doi/abs/10.1126/science.abg6116}{Science \textbf{375}, 193 (2021)}.

\bibitem{Dutta2022} B. Dutta, V. Umansky, M. Banerjee, and M. Heiblum, Isolated ballistic non-Abelian interface channel,
 \href{https://www.science.org/doi/10.1126/science.abm6571}{Science \textbf{377}, 1198 (2022)}.

\bibitem{Chand1990}
N. Chand and S.N.G Chu, A comprehensive study and methods of elimination of oval defects in MBE-GaAs, \href{https://doi.org/10.1016/0022-0248(90)90151-A}{J. Crystal Growth \textbf{104}, 485 (1990)}.

\bibitem{haug2019} 
Y. J, Chung, K. W. Baldwin, K. W. West, N. Haug, J. van de Wetering, M. Shayegan, and L. N. Pfeiffer, Spatial mapping of local density variations in two-dimensional electron systems using scanning photoluminescence,
\href{https://doi.org/10.1021/acs.nanolett.8b05047}{Nano Lett. \textbf{19}, 1908 (2019).} 


\bibitem{Yacoby2005} O. M. Auslaender, H. Steinberg, A. Yacoby, Y. Tserkovnyak, B. I. Halperin, K. W. Baldwin, L. N. Pfeiffer, and K. W. West, Spin-charge separation and localization in pne dimension, 
\href{https://doi.org/10.1126/science.1107821}{Science \textbf{308}, 88 (2005)}.

\bibitem{Rezayi2009} E. H. Rezayi and N. Read, Non-Abelian quantized Hall states of electrons at filling factors 12/5 and 13/5 in the first excited Landau level, \href{https://doi.org/10.1103/PhysRevB.79.075306}{Phys. Rev. B \textbf{79}, 075306 (2009)}.




\end{thebibliography}
\end{document}